\documentclass[prl,twocolumn,showpacs,showkeys]{revtex4}

\usepackage{graphicx}
\usepackage{dcolumn}
\usepackage{bm}
\usepackage{mathrsfs}

\usepackage{epsfig}

   \usepackage{amssymb}
   \usepackage{float}

\newcommand{\D}{\mathrm{d}}
\newcommand{\DD}{\mathrm{D}}

\newcommand{\Exp}[1]{\text{e}^{#1}}
\newcommand{\binomial}[2]{\left(\begin{array}{c} #1 \\ #2 \end{array} \right)}

\begin{document}
\title{Relativistic Fluctuation Theorems}

\author{A. Fingerle}
\email{axel.fingerle@ds.mpg.de}
\affiliation{%
Max Planck Institute for Dynamics and Self-Organization\\
Bunsenstr. 10, Germany - 37073 Goettingen
}%

\date{\today}

\begin{abstract}
Fluctuation Theorems are statements about the entropy of systems
far from thermal equilibrium. In this Letter relativistic
Fluctuation Theorems for Brownian motion are presented and proven.
Though there is a known discretization dilemma leading to a
certain class of Brownian processes, the Fluctuation Theorems are
shown to cover the entire class, and to
determine the physical process within this class. In the
general-relativistic case the heat responsible for entropy
production is the exchanged 4-momentum projected on the local time
axis of the heat bath.
\end{abstract}

 \maketitle

\emph{Introduction.} --- The physical basis of the direction of
time is being discussed since Boltzmann's H-Theorem in 1872. A
priori, this thermodynamic arrow of time has to be distinguished
from the possibility of a prime direction of time defined by the
monotonous expansion of our Universe \cite{Zeh}. Though we know by
now that due to its dominating Dark Energy component our Universe
is very likely to expand forever, a direct connection between
these arrows of time seems unlikely \cite{Hawking1, Page,
Hawking2}, leaving the fascinating cosmological arrow of time in a
speculative stage \cite{Spec1, Spec2}. Highly advanced is the
situation of the arrow of time in thermodynamics. Since 1993
Fluctuation Theorems (FTs) have been derived for very general
classes of systems \cite{Evans, GallavottiCohen1,
GallavottiCohen2, Kurchan, Jarzynski1, Jarzynski2, Udo}. Not only are these
FTs capable of proving the second law of thermodynamics for the
entropy $\Delta S$ produced by an ensemble,
\begin{eqnarray}
  \Delta S \ge 0 \text{ at any time,} \label{SL}
\end{eqnarray}
they even allow statements about the probability of ``violations''
of (\ref{SL}) by a single member of the ensemble in finite regions
of space and time. For the steady state of sufficiently chaotic
dynamical systems, FTs of the form
\begin{eqnarray}
  \frac{\text{P}(\Delta s=+a)}{\text{P}(\Delta s=-a)} = \exp
  \frac{a}{k_{\text{B}}}, \label{FT1}
\end{eqnarray}
valid for any $a$ have been proven \cite{Evans, GallavottiCohen1,
GallavottiCohen2}. Here $\text{P}(\Delta s=\nolinebreak a)$ is the
probability to observe a production of entropy equal to $a$ along
a trajectory segment. In the context of dynamical systems, the
dimensionless entropy ${\Delta s}/{k_{\text{B}}}$ is defined by
the contraction of phase space. The Steady State FT (\ref{FT1})
was derived in \cite{Kurchan} for a stochastic systems in contact
with a bath at temperature $T$. For this stochastic systems the
entropy $\Delta s = \Delta Q / T$ is caused by an energy transfer
$\Delta Q$ into the bath.
In \cite{Jarzynski1, Jarzynski2, Udo} a certain entropy production
$\Delta s$ is averaged exponentially to find Integral FTs of the
form
\begin{eqnarray}
  \left< \Exp{-\Delta s/k_{\text{B}}} \right> = 1, \label{FT2}
\end{eqnarray}
which remain valid even for non-stationary states in the presence
of time-dependent external forces. Angle brackets denote averaging
over trajectories by path integration \cite{Kurchan}.
Equation (\ref{FT1}) implies (\ref{FT2}), and Eq.~(\ref{FT2})
implies (\ref{SL}) by virtue of the Jensen inequality using
$\Delta S = \left<\Delta s\right>$.

Recently, the 1905 publications of Einstein on Brownian Motion
\cite{Einstein1} and Special Relativity \cite{Einstein2} were
unified \cite{Debbasch, DunkelHaenggi1, DunkelHaenggi2}. Based on
that, the results of this Letter are threefold. First, for the
relativistic Brownian processes \cite{DunkelHaenggi1} and
\cite{DunkelHaenggi2}, we reconcile the FTs (\ref{FT1}) and
(\ref{FT2}), which have become a paradigm of nonequilibrium
physics, with Special Relativity. FTs for a similar 
process put forward in \cite{Debbasch} can be derived readily by
an analogous thread of reasoning.

In \cite{DunkelHaenggi1} and \cite{DunkelHaenggi2} it was pointed
out that the relativistic time dilation leads to multiplicative
coupling, necessitating a careful choice of the three
discretization rules (according to It\^o, Fisk-Stratonovich, as
well as H\"anggi-Klimontovich), since they lead to physically
different processes. As second central result we show, that there
is one relativistic Steady State FT (\ref{FT1}) and one
relativistic Integral FT (\ref{FT2}) valid for all choices.
Furthermore, we shall find the physically correct expression for
the entropy production following from relativistic FTs
when the H\"anggi-Klimontovich discretization rule is applied.

The third result of this Letter is a general-relativistic Integral
FT for the Cosmological Standard Model. 
This exposes clearly the role of cosmic expansion in
entropy production. We shall identify the entropy production which
is solely due to the Hubble expansion of space. Such entropy
producing processes dominate when the expansion rate of the
Universe exceeds the particle scattering rate, for instance in an
early inflationary phase after the Big Bang.


\emph{Relativistic Brownian Motion.} --- To avoid technicalities
we consider first the one-dimensional motion of a test particle in
a heat bath.
The generalization to higher spatial dimensions is
straightforward. In the high temperature limit, the mean squared
velocity of this particle with rest mass $m$ may no longer obey
the nonrelativistic law $\overline{v^2}=k_{\text{B}} T/m$ since
the finite speed of light defines an insurmountable upper bound.
The authors of \cite{Debbasch, DunkelHaenggi1, DunkelHaenggi2}
have set forth the
relativistic Brownian motion giving rise to relativistic velocity
distributions in both, the language of stochastic differential
equations (relativistic Langevin equations) and the language of
probability densities (relativistic Fokker-Planck equations).
Following \cite{DunkelHaenggi1, DunkelHaenggi2}, the generalized
deterministic force $F_{\text{d}}$ in the rest frame of the heat
bath is
\begin{eqnarray}
\D p_{\text{d}} = F_{\text{d}} \ \D t =-\nu p \label{Fdet} \ \D t
\ ,
\end{eqnarray}
so that the time scale of dissipation is $1/\nu$.
Equation~(\ref{Fdet}) is of the familiar form \cite{Risken} with
the nonrelativisic momentum $mv$ substituted by
$p=p^1={mv}/{\sqrt{1-{v^2}/{c^2}}}$, which is the spatial
component of the relativistic momentum vector $p^\alpha$. As is
common, Greek indices refer to temporal ($\alpha=0$) and spatial
components. The signature of the Minkowski metric tensor is
$\eta_{\alpha \beta}=\eta^{\alpha \beta}=\text{diag}(-1,1)$.
Moreover, Einstein's summation convention is invoked throughout.
Since the rest mass is not changed in collisions, $p^\alpha
p_\alpha=-(mc)^2=\text{const}$, the change in the momentum vector
$\D p^\alpha$ is always ``orthogonal'' to $p_\alpha$ in the sense
of
\begin{eqnarray}
p_\alpha \D p^\alpha=0 \ . \label{orth}
\end{eqnarray}
This means that the classical particle cannot leave its mass shell
$p^\alpha p_\alpha=-(mc)^2$, which is nothing but its dispersion
relation,
\begin{eqnarray}
E=p^0 c=\sqrt{(mc^2)^2+(pc)^2} \ . \label{DispRel}
\end{eqnarray}
The general solution of (\ref{orth}) is the projection
\begin{eqnarray*}
\D p^\alpha= \left(\delta^\alpha_\beta + \frac{p^\alpha
p_\beta}{(mc)^2}\right) \xi^\beta
\end{eqnarray*}
of an arbitrary Lorentz vector $\xi^\beta$. It is readily seen
that Eq.~(\ref{Fdet}), valid in the rest frame of the bath,
implies
\begin{eqnarray}
\D p_{\text{d}}^\alpha= -m \nu \left(\delta^\alpha_\beta +
\frac{p^\alpha p_\beta}{(mc)^2}\right) v_{\text{bath}}^\beta \D
\tau \ , \label{ViscTens}
\end{eqnarray}
with the bath velocity vector $v_{\text{bath}}^\beta$ and the
particle's proper time $\tau$.
Equation~(\ref{ViscTens}) is the generalized Lorentz-invariant
deterministic part of the Brownian motion \footnote{Equation~(16)
in \cite{DunkelHaenggi1} contains a vanishing
term.}.

The description of relativistic Brownian motion is completed by
Lorentz-invariant stochastic changes $\D p^\alpha_{\text{s}}$ of
the momentum caused by the impacts of the surrounding heat bath at
temperature $T$. The derivation in \cite{DunkelHaenggi1,
DunkelHaenggi2} rests on two ideas: The relativistic momentum is
the proper quantity performing a Wiener process, since it is
physically exchanged and additive, whereas the velocity is
well-known to be not additive in Special Relativity.
The second postulate demands that the distribution is a Gaussian
in the instantaneous rest frame of the particle. This connects the
relativistic Brownian motion to the nonrelativistic case. These
principles determine the exchanged momenta $\D
p^\alpha_{\text{s}}$ do be distributed according to (cf. Eq.~(35c)
in \cite{DunkelHaenggi1})
\begin{eqnarray}
P_{\text{coll}}(p^\mu,\D p^\nu_{\text{s}})
= \frac{mc \ \delta\left(p_\beta
\D p^\beta_{\text{s}}\right)}{2\sqrt{\pi {\mathscr D} \D \tau}}
\exp -\frac{\D p^\alpha_{\text{s}}\D p_{\text{s} \; \alpha}}{4 {\mathscr D}
\D \tau}. \label{1u1Verteilung}
\end{eqnarray}
The Dirac distribution $\delta(p_\beta \D p^\beta_{\text{s}})$
guarantees that the mass-shell condition (\ref{orth}) is also
fulfilled by the stochastic impacts, since they are elastic.
While the relativistic momentum $p$ is additive and
unbounded, the velocity
is restricted to the open interval $(-c,+c)$. This can be seen by
the elegant relation $v/c^2=p/E$ in the rest frame of the bath,
which is equivalent to
\begin{eqnarray}
\D x = \frac{pc}{\sqrt{(mc)^2+p^2}} \ \D t \ . \label{Ortsanteil}
\end{eqnarray}

The bath temperature $T$ is defined
by the Einstein relation for the momentum diffusion constant
${\mathscr D}$ (cf. Eq.~(59) in \cite{DunkelHaenggi1}):
\begin{eqnarray}
{\mathscr D} = k_{\text{B}}T m \nu \ . \label{DiffConst}
\end{eqnarray}
The Einstein relation is known to apply far from equilibrium
\cite{Udo}. This is not surprising since (\ref{DiffConst}) relates
the bath temperature to the strength of stochastic impacts
${\mathscr D}$, and the particle damping rate $\nu$. These
quantities define the coupling between the bath and a single
particle, be it considered as part of an equilibrium ensemble or
not.

\emph{Relativistic Fluctuation Theorem.} --- We have now the
manifestly Lorentz-invariant Langevin equation
\begin{eqnarray}
\D p^\alpha = \D p_{\text{d}}^\alpha + \D p_{\text{s}}^\alpha
\label{LILE}
\end{eqnarray}
with the deterministic part given by (\ref{ViscTens}) and the
stochastic part described by (\ref{1u1Verteilung}) at hand.
Specifying (\ref{LILE}) to the rest frame of the bath,
$v_{\text{bath}}^\alpha=(c,0)$, yields
\begin{eqnarray}
\D p = - \nu p \ \D t + \D p_{\text{s}} \label{BathLE} \ .
\end{eqnarray}
The probability density of the exchanged momenta $\D p_{\text{s}}$
is found by integrating out the $\D p^0_{\text{s}}$-component in
(\ref{1u1Verteilung}), cf.~\cite{DunkelHaenggi1}:
\begin{eqnarray}
P_{\text{coll}}(p,\D p_{\text{s}}) = \frac{\exp{\left(-\D p_{\text{s}}^2 /4
{\mathscr D} \sqrt{1+\frac{p^2}{(mc)^2}} \D t\right)}}{2\sqrt{\pi
{\mathscr D} \D t}\sqrt[4]{1+\frac{p^2}{(mc)^2}}} \ .
\label{1Verteilung}
\end{eqnarray}
The Eqs.~(\ref{Ortsanteil}), (\ref{BathLE}) and
(\ref{1Verteilung}) establish the relativistic stochastic motion
of the Brownian particle in phase space. The corresponding
transition probability is easily determined by the discretization
rule, for example in the case of It\^o:
\begin{eqnarray}
P_{\text{trans}}\binomial{x\mapsto x+\D x}{p \mapsto p+\D p} = \frac{\delta \left(\D x - \frac{pc \ \D
t}{\sqrt{(mc)^2+p^2}}\right)}{2\sqrt{\pi {\mathscr D} \D
t}\sqrt[4]{1+\frac{p^2}{(mc)^2}}} \nonumber \\ \times \exp{-\frac{
\left( \D p + \nu p \D t\right)^2}{4 {\mathscr D}
\sqrt{1+\frac{p^2}{(mc)^2}}\D t}} \ . \label{PSVerteilung}
\end{eqnarray}

It has been clarified in \cite{Udo} that entropy fluctuations are
caused by the particle entropy $s_{\text{p}}=\nolinebreak
-k_{\text{B}} \ln P({\bf x}(t),{\bf p}(t),t)$ with the particle's
nonequilibrium phase space density $P({\bf x},{\bf p},t)$, and the
entropy $s_{\text{m}}$ of the surrounding medium at temperature
$T$. From the equation of motion (cf. Eq.~(7) in \cite{Udo}) for
$s_{\text{p}}$,
\begin{eqnarray}
\D s_{\text{p}}=\D s_{\text{p}}\vert_{\lambda=0} + \lambda
k_{\text{B}} \ \D \ln E \ , \label{dsp}
\end{eqnarray}
we have isolated the second term which depends on the
discretization rule applied to (\ref{BathLE}):
H\"anggi-Klimontovich, Fisk-Stratonovich, or It\^o correspond to
$\lambda=\nolinebreak 0,$ $\frac{1}{2}$, or $1$ respectively.

The entropy production $\D s_{\text{m}}$ in the bath follows by
contrasting the probability of a trajectory $(x,p)^{t_f}_{t_i}$
with its time-reverse
$(x,p)^\dagger(t)=\left(x(t_f-t),-p(t_f-t)\right)$ to extract the
time-asymmetric part causing dissipation (cf.~\cite{Udo} and
references therein):
\begin{eqnarray}
\D s_{\text{m}} \equiv k_{\text{B}} \ln
\frac{P_{\text{trans}}}{P_{\text{trans}}^\dagger}= -\frac{\D E}{T}
- \lambda k_{\text{B}} \ \D \ln E \ . \label{dsm}
\end{eqnarray}

From the Eqs.~(\ref{dsp}) and (\ref{dsm}) we see that though the
relativistic motion of the Brownian particle is physically
inequivalent depending on $\lambda$, the fluctuations of the total
entropy $s=s_{\text{p}}+s_{\text{m}}$ are independent of
$\lambda$. The Eqs.~(\ref{dsp}) and (\ref{dsm}) can readily be
integrated by the method developed in \cite{Udo} over a finite
time interval to the Steady State FT (\ref{FT1}) and for
time-depend states
to the Integral FT (\ref{FT2}). Therewith we have proven
relativistic FTs that are unaffected by the discretization
dilemma.


We are now in the position to address the physical choice of
$\lambda$ by virtue of the FT. In the nonrelativistic regime, $E$
is dominated by the rest mass so that the second term in
(\ref{dsm}) vanishes for all $\lambda$. At arbitrary relativistic
energies (\ref{DispRel}) the H\"anggi-Klimontovich rule,
$\lambda=0$, yields the correct expression for the entropy
\begin{eqnarray}
\D s_{\text{m}} =  -\frac{\D E}{T} \ , \label{result1}
\end{eqnarray}
which is produced in the heat bath.

\emph{Generalizations in the framework of Special Relativity.} ---
To generalize the FTs to $n$ spatial dimensions, momentum and
force in Eq.~(\ref{Fdet})
are simply substituted by their spatial vectors and the Greek
indices in the Lorentz-invariant Eqs.~(\ref{ViscTens}) and
(\ref{1u1Verteilung}) take values up to $n$. After integrating out
the temporal component $p^0$, the distribution (\ref{1Verteilung})
is found to contain a quadratic form
$\bf A$ instead of the square in the exponent (cf. Eq.~(15) in
\cite{DunkelHaenggi2}) with tensor components
\begin{eqnarray}
A_{ij}= \ \delta_{ij} - \frac{c^2}{E^2} p_i p_j \ . \label{Amarix}
\begin{array}{ccc}
\end{array}
\end{eqnarray}
The FTs follow using the obvious fact that $\bf p$ is eigenvector
of $\bf A$.

No complications are caused by allowing an inhomogeneous heat
bath, where the dissipation rate $\nu$ is a function of space and
time. The dissipation rate $\nu$ may also be an even function of
the momentum, $\nu(\vert p \vert)$. This of importance since we
may not expect the embedding medium to behave as a Newtonian fluid
at relativistic energies. A temperature varying in space and time
does not pose a problem, nor does an additional time-dependent
external force.
In this case the produced heat $\D Q=T \D s_{\text{m}}$
is the loss of particle energy in collisions, $-\D E$ reduced by
the work extracted via the external force.


\emph{Generalizations in the framework of General Relativity.} ---
The monotonic increase of entropy is a fundamental principle of
physics
and the Universe is known to expand, as was discovered by E.
Hubble in 1929. The discussion whether there is a direct
connection between these observations has newer stopped
\cite{Hawking1, Page, Hawking2, Spec1, Spec2}. Therefore we aspire
a formulation of the FT consistent with General Relativity, but we
restrict ourselves to the class of Friedmann-Lema\^itre models,
which describe a spatially homogenous and isotropic, expanding or
contracting Universe. The corresponding line element (as given by
the Robertson-Walker metric) is $-\D t^2 + \D r^2$. The important
difference compared to Special Relativity is that the spatial
part, $\D r^2$, is scaled by a time dependent factor $R(t)$
describing the expansion or contraction of the Universe:
\begin{eqnarray}
\D r^2 = R^2(t) \ h_{ij}(\xi) \ \D \xi^i \ \D \xi^j \ .
\end{eqnarray}
The Latin indices describe spatial components numbered by $1$ to
$n$. We do not have to deal with the details of the metric tensor
$\bf h$ describing the spatial geometry. The result will be valid
for arbitrary geometries.
The expansion rate $H(t)=\dot R(t)/R(t)$, named Hubble function,
is one of the most important quantities in cosmology and its
present value is a direct observable \cite{Peacock}. The typical
frame for a cosmic heat bath is the frame of the cosmic microwave
background.

\begin{figure} \begin{center}
\epsffile{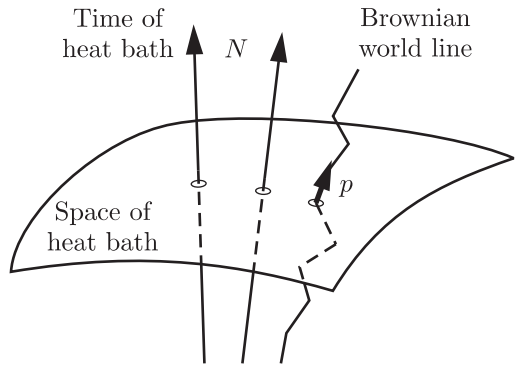} \caption{A sketch of spacetime showing a
spatial slice of the heat bath at fixed time and the world line of
a Brownian particle in a (locally) expanding Universe.}
\label{Graphic}
\end{center} \end{figure}

In General Relativity, the correct equations of motion include the
covariant differential $\DD p$ of the momentum.
(Denoting by $p$ the 4-vector, the components of $\DD p$ are $\DD
p^\alpha=\D p^\alpha + \Gamma_{\mu \nu}^\alpha p^\mu \D x^\nu $.)
Its spatial components replace the left hand side of
(\ref{BathLE}) and can be split up into a spatially covariant
part, ${}^{(n)}\DD {\bf p}$, and a contribution due to the
time-dependent scaling:
\begin{eqnarray}
\DD {\bf p} = {}^{(n)}\DD {\bf p} + H(t) {\bf p} \ .
\end{eqnarray}
Therefore the covariant Langevin equation, generalizing
Eq.~(\ref{BathLE}) to be valid in an expanding or contracting
Universe of arbitrary spatial geometry, reads
\begin{eqnarray}
{}^{(n)} \DD {\bf p} =  - \left[\nu(\Vert {\bf p}\Vert,t) +
H(t)\right] {\bf p} \ \D t + {}^{(n)} \DD {\bf p}_{\text{s}} \ .
\end{eqnarray}
The distribution of the stochastic impacts ${}^{(n)} \DD {\bf
p}_{\text{s}}$ is found after substituting the Euclidean metric
$\delta_{ij}$ by $h_{ij}$ in (\ref{Amarix}). 
Applying the time-reversal map we find that Eq.~(\ref{result1})
gains a second term:
\begin{eqnarray}
\D s_{\text{m}}
&=& -\frac{\D E}{T}
\ - \frac{\Vert{\bf p} c\Vert^2}{E T} \ \D \ln R \nonumber \\ 
&=& \D s_{\text{static}} - H \frac{\left({\bf p},\D {\bf
r}\right)}{T} \ . \label{result2}
\end{eqnarray}
The numerator $\left({\bf p},\D {\bf r}\right)$ in (\ref{result2})
is the canonical line integral (canonical one-form) in phase
space. The Integral FT (\ref{FT2}) extends to an expanding ($H>0$)
or contracting ($H<0$) spacetime when this second term is taken
into account. It has a clear geometric interpretation: The Hubble
function is the external curvature of space,
\begin{eqnarray}
\DD N = H \ \D {\bf r}\ , \label{DivConvergence}
\end{eqnarray}
with $N$ being the time-like normal vector to the space of the
heat bath as depicted in Fig.~\ref{Graphic}. This permits the
second term in (\ref{result2}) to be written as
\begin{eqnarray*}
-\frac{(p,\DD N)}{T} \ .
\end{eqnarray*}
Since the particle energy $E=p^0=-p_0=-(p,N)$ is the zero
component of the 4-vector $p$, the first term in (\ref{result2})
equals the differential
\begin{eqnarray*}
\frac{\D (p,N)}{T} = \frac{(\DD p,N)+(p,\DD N)}{T} \ ,
\end{eqnarray*}
such that the sum of both terms is the projection
\begin{eqnarray}
\D s_{\text{m}} = \frac{(\DD p,N)}{T} \ . \label{result3}
\end{eqnarray}
As described by Eq.~(\ref{DivConvergence}), the geodesic flow $N$
may be convergent or divergent, corresponding to a contracting or
expanding Universe, but this does not imply a change of sign for
the entropy production $\D s_{\text{m}}$.

\emph{Conclusions.} --- Relativistic FTs have been established
that remain valid in the regime of high temperatures or low
masses, $mc^2\ll k_{\text{B}} T$.
By investigating the entropy production for particle and
environment separately, we could determine the physically correct
discretization rule, which had not been possible sofar by
relativistic invariance alone. These FTs were found to extend in
the framework of General Relativity. Such a formulation reveals
the influence of a time-depended gravitation field on the local
entropy production. The dissipated heat $T \D s_{\text{m}}$ is the
exchanged 4-momentum $\DD p$ projected on the local time direction
of the heat bath.

On the theoretical road ahead, one may expect Integral FTs to hold
for arbitrary time-dependent and inhomogeneous fields, such as
gravitational waves, when the concise expression (\ref{result3})
is applied.
For the process originally introduced in \cite{Debbasch}, the
weaker inequality (\ref{SL}) has been proven recently
\cite{Debbasch2} under general conditions. Experimentally, the
relativistic FT is not only subject of high energy physics and
cosmology. An ultra relativistic FT can be tested with a
high-precision spectroscopy experiment by shining with a laser on
an exited granulate of glass beads, so that the granulate serves
as a heat bath and the photons are the relativistic ``Brownian''
particles.
\begin{acknowledgments}
I am grateful for comments of J.~Dunkel, S.~Herminghaus,
U.~Seifert and M.~Brinkmann . 
\end{acknowledgments}

\end{document}